# Particle Localization by Decoherence and Classical Lensing

*Ivo Knittel, Experimental Physics Department, Saarland University, Saarbrücken, Germany*

A phenomenological model of the time evolution of a particle wavepacket is presented that is subject to scattering event with small momentum transfer. It is suited for three dimensions and allows for an additional potential. For a random value of a phase parameter, it is equivalent to the decoherence by scattering off plane wave field states. Imposing some condition on the same phase parameter, the model delivers definite outcomes in a local pointer base. Common sense notions about the quantum-classical transition are readily obtained: Decoherent free particle solutions are localized constant-width wavepackets, quantum barrier scattering results in random reflection and transmission. The scattering process assumed in the phenomenological model is realized by scattering of a particle pair interacting by a force-distance law with small momentum transfer. Particles are localized not by decoherence but by a classical lensing effect. Schmidt paths in the Schmidt representation of the multiparticle state are products of single-particle wavepackets. Therefore the model has a natural 'Many Worlds' interpretation, identifying each Schmidt path with a classical branch. Different runs of the phenomenological model map the state of the same particular particle within different branches.

Erwin Schödinger hoped there may arise some intrinsic mechanism within wave mechanics resulting in constant-width wavepacket solutions to be identified with free classical particles.[1] It was not until the 80ies that some progress at this problem has been achieved: Within the theory of decoherence [2-4], plane wave decoherence models (PWD show that for massive particles interacting with a radiation field, the decoherence length can be many orders smaller than the momentum mean free path.[5-7] Assuming frequent scattering of field quanta by the particle, an equation of motion for the density matrix is derived. On this basis, Ref. 4 discusses in detail the propagation of a particle wavepacket under the influence of decoherence. It is found that the width of the probability density of an initial Gaussian wavepacket increases at the same rate as it would in the absence of decoherence, or more, if classical recoil effects are present. However, correlations among position measurements at different positions at the same time decay, governed by a finite decoherence length. Among decoherence models, PWD is applied with most success to experiments. [8-11]

The reduced density matrix of a particle is a complete record of measurement at the particle state alone. When the reduced density matrix is identical with that of an ensemble of wavepackets, then there is no physical difference between the mere loss of spatial correlation, and the presence of wavepackets.[12] On this basis, has been argued that the ensemble interpretation of the reduced density matrix is unnecessary for quantitative physical predictions. [12] However the necessity of a genuine model of the collapse of the wavefunction with a definite outcome has always been felt. Collapse models exist within the Relative State [13] and also are closely related Consistent Histories approaches[14-15] .

In multiparticle systems, particle positions can be correlated, so that fixing one particle in positions may lead to localization of all other particles in the system. Superpositions of correlated states are the base of Relative State interpretations of quantum mechanics. [16] A more formal structure that is widely used in this context ist the Schmidt representation - the overall state evolving in time can be written at any instant as a sum of products of one-particle wavefunctions. Any time-dependent term in the sum defines uniquely a Schmidt path. In Ref. 17, a wavefunction collapse is produced in a toy model within the interpretatory framework of Many Worlds and Consistent Histories. If a single Schmidt component can consists of localized one-particle states, and its amplitude is reasonably stabe over time, it cna be interpreted as a classical branche within a Relative State interpretation. However in general, Schmidt components do not resemble classical worlds. A review of the relations between decoherence and collapse models and the corresponding difficulties is given in Ref. 18. In PWD, there is no direct relation between the Schmidt components and the pointer states that can be found in the reduced density matrix, e.g. by applying the 'predictability sieve' [19]. Ref. 16 pointed to the limited scope of PWD and argued that in other real-world situations, may well be described decoherence models with all-classical Schmidt paths. In two-particle scattering, classical Schmidt paths were found in a limited parameter range. [20] Another direction within Relative State concepts does not rely on Schmidt paths, but on robustness criteria.[21-23] However also these approaches suffer from their own difficulties related to the Born rule. [18] In this situation, it is of interest to discuss special models that do not suffer from the difficulties that arise in the general case.

In the first part of this work, a phenomenological model of small-momentum scattering is developed. Numerical results on some standard situations are shown. In the second part, the relation of the phenomenological model to fundamental quantum theory is discussed. For that purpose, we discuss the entanglement states arising from two-particle scattering. On that basis, the Schmidt decomposition of a gas of particles after successive scattering events is discussed.

**Phenomenological decoherence model**

In order to develop a model that produces particle localization by decoherence (PLD), we assume the coherent dynamics of a particle wavepacket $\psi(\boldsymbol{x},t)$, to be interrupted at times $\{t_{j}, j = 1 \ldots \}$ by events. At an event, a perturbed wavefunction is produced that is shifted slightly in momentum,

$$\psi(\mathbf{x}) \rightarrow e^{i\kappa\cdot(\mathbf{x}-\mathbf{x}_0)}\psi(\mathbf{x}) \tag{1}$$

using a WKB approximation, which is appropriate for a quasi-free particle. The position of zero phase $\boldsymbol{x}_0$ is chosen randomly, but proportional to the probability density. It can be regarded as the position of the scattering event. The perturbed wavefunction is added to the unperturbed wavefunction with some amplitude,

$$\psi(\mathbf{x},t+0) = \psi(\mathbf{x},t-0) + g\; e^{i\,\kappa\cdot(\mathbf{x}-\mathbf{x}_0)}\;\psi(\mathbf{x},t-0) \tag{2}$$

with $g = \gamma e^{i\varphi}$. Amplitude and relative phase of the shifted wavefunction are denoted by a nonnegative real $\gamma$ scattering amplitude and the relative phase $\varphi$ at the scattering position $\boldsymbol{x}_0$. The momentum transfer is $\boldsymbol{\kappa} = \xi\boldsymbol{\kappa}_0\,\boldsymbol{\zeta}$, with a random unit vector $\boldsymbol{\zeta}$, and a random number $\xi\in[-1/2,1/2]$. The phase $\varphi$ will turn out to be essential.
The limit of small momentum transfer $\boldsymbol{\kappa}$ is of particular interest, when the width of the wavepacket is much smaller than the momentum transfer wavenumber

$$\sigma << 2\pi/\boldsymbol{\kappa}\,. \tag{2b}$$

Then, one parameter can be eliminated: Then, the typical phase shift can be approximated by

$$\psi(\mathbf{x},t+0) = \psi(\mathbf{x},t-0) + \frac{ig\kappa}{(1+g)}(x-x_0)\;\psi(\mathbf{x},t-0). \tag{3}$$

The same equation is however obtained for other parameter values, too, as long as holds

$$\frac{(1+\tilde{g})g}{(1+g)\tilde{g}} = \frac{\tilde{\kappa}}{\kappa}. \tag{4}$$

In the recoil-free limit (3), one can always assume $\tilde{\gamma}$ =1/2, adjusting the value for the momentum transfer to

$$\tilde{\kappa} = \frac{3\gamma}{(1+\gamma)}\kappa., \quad \tilde{\kappa} = \frac{\gamma}{(1-\gamma)}\kappa., \tag{5}$$

For φ = 0 , and for the other important case φ = π, respectively.

**Implementation**

The approach outlined above is applicable to three dimensional problems. It is demonstrated in the following in one dimension. For the coherent wavepacket dynamics we use an explicit method [24-26]. The time-dependent Schrödinger equation

$$i\partial_t \psi(x,t) = -\frac{1}{2m}\partial_x^2 \psi(x,t) + V(x)\psi(x,t) \tag{6}$$

is written in terms of real functions,

$$\begin{aligned}
&\psi(x,t) = R(x,t) + iI(x,t), \\
&\partial_t R(x,t) = -\frac{1}{2m}\partial_x^2 I(x,t) + V(x)I(x,t), \\
&\partial_t I(x,t) = +\frac{1}{2m}\partial_x^2 R(x,t) - V(x)R(x,t).
\end{aligned} \tag{7}$$

A Taylor expansion of R and I leads to

$$\begin{aligned}
&R(x,t+\frac{\tau}{2}) \cong R(x,t-\frac{\tau}{2}) - 2\{\alpha[I(x+\xi,t) + I(x-\xi,t)] - 2[\alpha + V(x)\tau]I(x,t)\}, \\
&I(x,t+\frac{\tau}{2}) \cong R(x,t-\frac{\tau}{2}) + 2\{\alpha[R(x+\xi,t) + R(x-\xi,t)] - 2[\alpha + V(x)\tau]R(x,t)\}, \\
&\alpha = \frac{\tau}{2\xi^2}.
\end{aligned} \tag{8}$$

In discrete form, this becomes

$$\begin{aligned}
R_i^{n+1} &= R_i^n - 2\{\alpha[I_{i+1}^n + I_{i-1}^n] - 2[\alpha + V_i\tau]I_i^n\}, \\
I_i^{n+1} &= I_i^n + 2\{\alpha[R_{i+1}^n + R_{i-1}^n] - 2[\alpha + V_i\tau]R_i^n\},
\end{aligned} \tag{9}$$

$t=n\tau$, $x=i\xi$.

From the above algorithm we obtain the coherent time propagation

$$\psi(x,t+t_c) = \mathbf{U}_{t_c}\psi(x,t)\,, \tag{10}$$

The central part of the computation is the scattering event according to Eq. 2.

$$\psi(x,t-0) \mapsto \psi(x,t+0)$$

The parameters related to the decoherent process are the time between scattering events $t_c$, the mean momentum transfer $\kappa_0$, and the scattering amplitude, $\gamma$. We start with the relative phase $\varphi = 0$. The momentum shifts are box distributed between $-\kappa_0/2$ and $\kappa_0/2$. To obtain the scattering position $\mathbf{x}_0 = x_0$, we draw a box around the plot of the probability density $\rho(x)$ and shoot into it. That is, we generate random number pairs

$$(\, x_p \in [\, x_{min}, x_{max} \,],\ p_r \in [\, 0, \max(\, \rho(x)\,)\,]\,)\,. \tag{11}$$

$x_0$ is given by the first value $x_p$ we find for which $p_r < \rho(x_p)$. Normalizing the result, the wavefunction after the event $\psi(x,t+0)$ is obtained. The initial state is a Gaussian wavepacket of width $\sigma$ and position $x_{ini}$. The total simulation time is t. As the scattering event is random, the time evolution results in a member from an ensemble of wavefunctions. The ensemble size is denoted by $m$.

**Numerical results on particle localization by decoherence**

The time evolution of a Gaussian wavepacket in the presence and absence of decoherence is computed for $n$ = 750, d$x$ = 0.02, d$t$ = d$x^2$/2, and parameters $k_0 = 2.5\pi$; $t_c = 2dt$; $\sigma = 1.5$; $x_{ini} = 7.5$; $\kappa_0 = k_0/30$; $\gamma = \kappa_0/4$, which is equivalent to $\gamma$=1/2, $\kappa_0$=1/180. Results are shown for simulation time $t$ = 0.4 in Fig. 1. In Figs. 1(a) and 1(b) is demonstrated that decoherent propagation leads to a pronounced sharpening of the wavepacket, counteracting the familiar 'spreading of the wavefunction'. In Fig. 1c is displayed the density for the ensemble of wavefunctions after decoherent evolution. Interestingly, the ensemble density is also narrower than the density resulting from coherent evolution. In Fig. 1d, decoherent wavefunction densities after decoherent propagation are shown centered, i. e. with their maxima at the same position. Individual decoherent wavefunctions are similar, i.e. Gaussians of similar width – the random process results in a deterministic time evolution of the waveform. In 1(e) individual wavefunctions of the decoherent ensemble are displayed at their actual positions, compared with the wavefunction after coherent evolution on the right. The time evolution of the position and variance of the density of an individual decoherent wavefunction is displayed in Fig. 2. A superposition of two wavepackets

$$\psi(x,t=0) = \sqrt{\beta}\psi(x-d-2,0) + \sqrt{1-\beta}\psi(x+d/2,0) \tag{9}$$

is not stable under decoherent propagation, but decays in a random manner into one of the component wavepackets. This is demonstrated in Figs. 3 and 4. The conversion to a single wavepacket takes place at a much shorter timescale than the spread of the wavefunction investigated in the previous section. The model also reproduces the random character of the scattering of a quantum particle at a potential barrier. This case is shown in in Fig. 5.

**Localizing, delocalizing and neutral decoherent propagation**

In Eq. 2 defining the scattering event, we denoted by φ the relative phase between the scattered wavefunction and the original wavefunction at the scattering position $\boldsymbol{x}_0$. The value of this phase is crucial to the decoherent propagation. $\boldsymbol{x}_0$ is typically a position where the probabiltiy density is highest. At that position interference between the scattered and original wavefunction is constructive for $\varphi = 0$, and destructive for $\varphi = \pi$. The amplitude at the

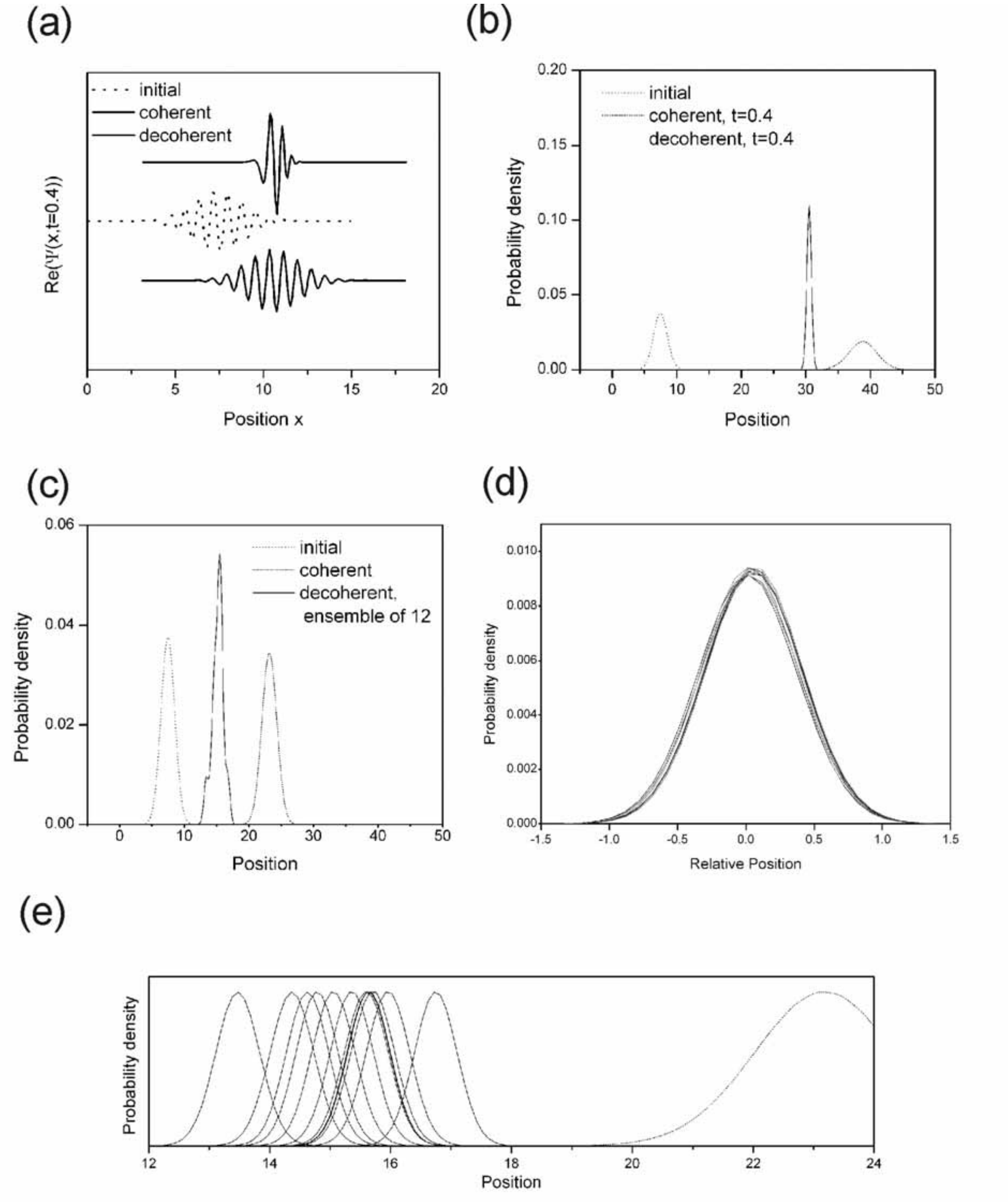

Fig. 1 (a): Re (Ψ) of the initial wavepacket (dotted), after coherent evolution (dashed), and decoherent evolution (full), simulation time t=0.4. (b): Probability densities, t=1. (c): the same as (b), but with the fill line indicating the probability density of an ensemble of 12 runs of decoherent evolution. (d): decoherent wavefunctions of the ensemble, compared with the wavefunction of the coherent evolution, all centered. (e): individual decoherent wavefunctions, compared with the coherent wavefunction, at their actual positions.

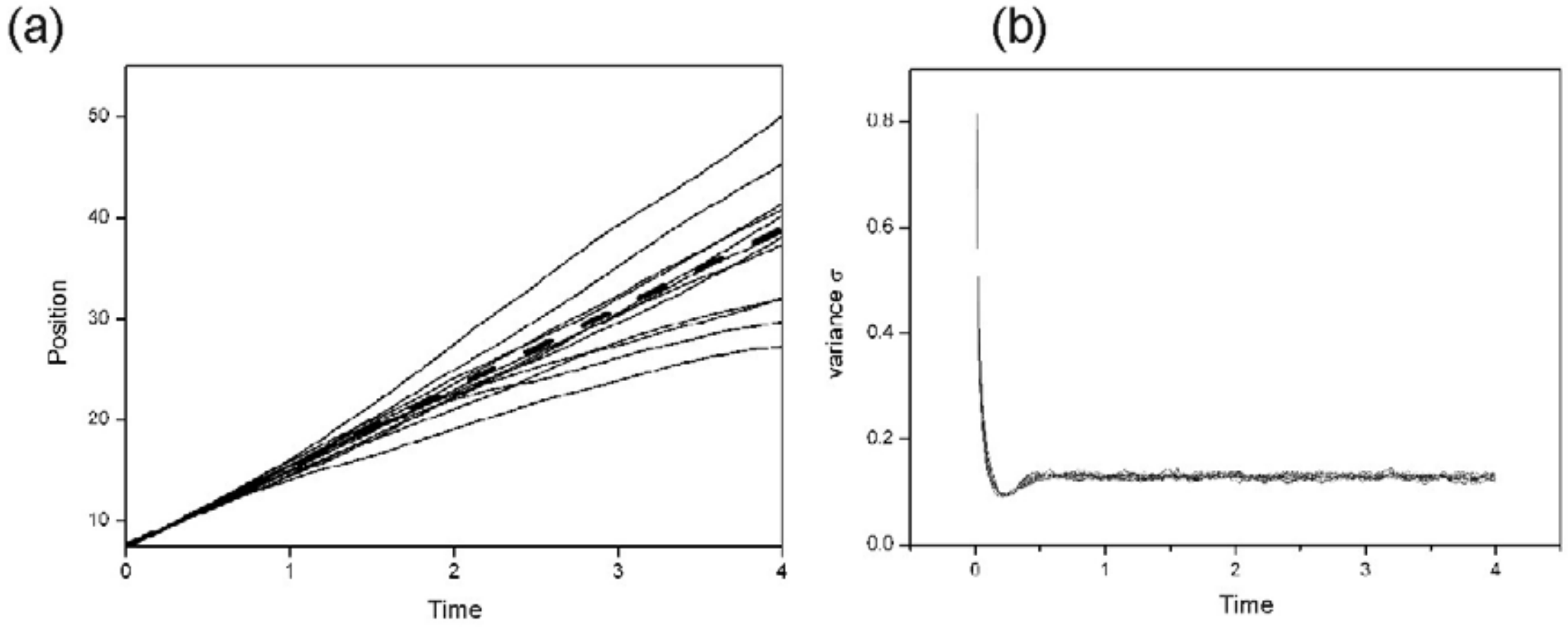

Fig. 2: (a) position of the wave-function maximum (circles) vs time, for 12 decoherent runs (full lines), and coherent propagation, $2k_0t$ (dashed). (b): the squared variance of position vs time of 12 decoherent wavepackets.

Fig. 3: (a) Initial double wavepacket state wavepacket (blue), and state after coherent (green) and decoherent (red) propagation;

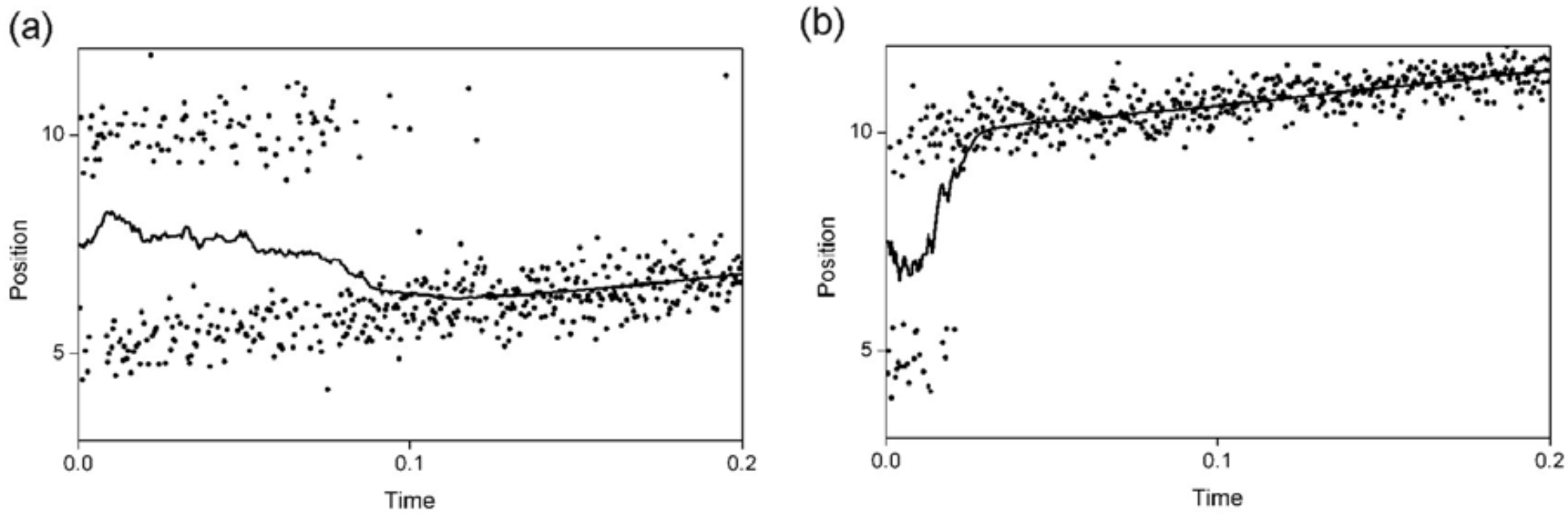

Fig. 4:Time evolution of the above double packet state. Line: expectation value vs time, dots indicate positions of scattering events, (a) with parameters (a) $\kappa_0 = k_o/30$, $\gamma = \kappa/4$, (b) $\kappa_0 = k_o/60$, $\gamma = \kappa_0/6$.

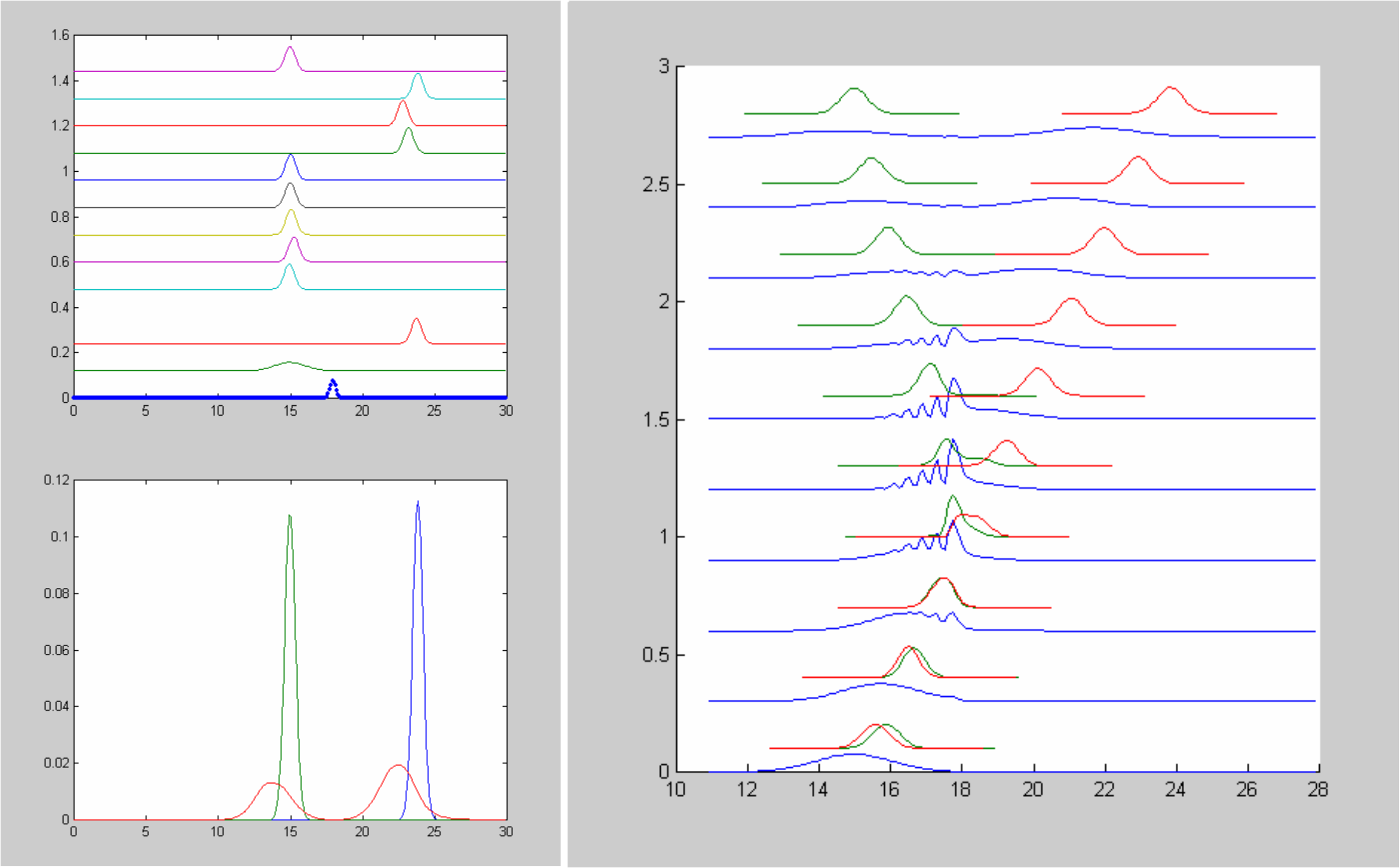

Fig. 5: Upper Left; Scattering of the initial wavepacket (density displayed green, 2nd from below) by the potential barrier (blue, below) in the presence of decoherent interaction. Any other line indicates a wavepacket density for a different run of the simulation, after time $t$=1. Lower Left; Wavepacket densities after $t$=1, after decoherent propagation, barrier transmission (blue), barrier reflection (green), and coherent propagation (red). Right: Wavepacket densities at time intervals of 0.1.

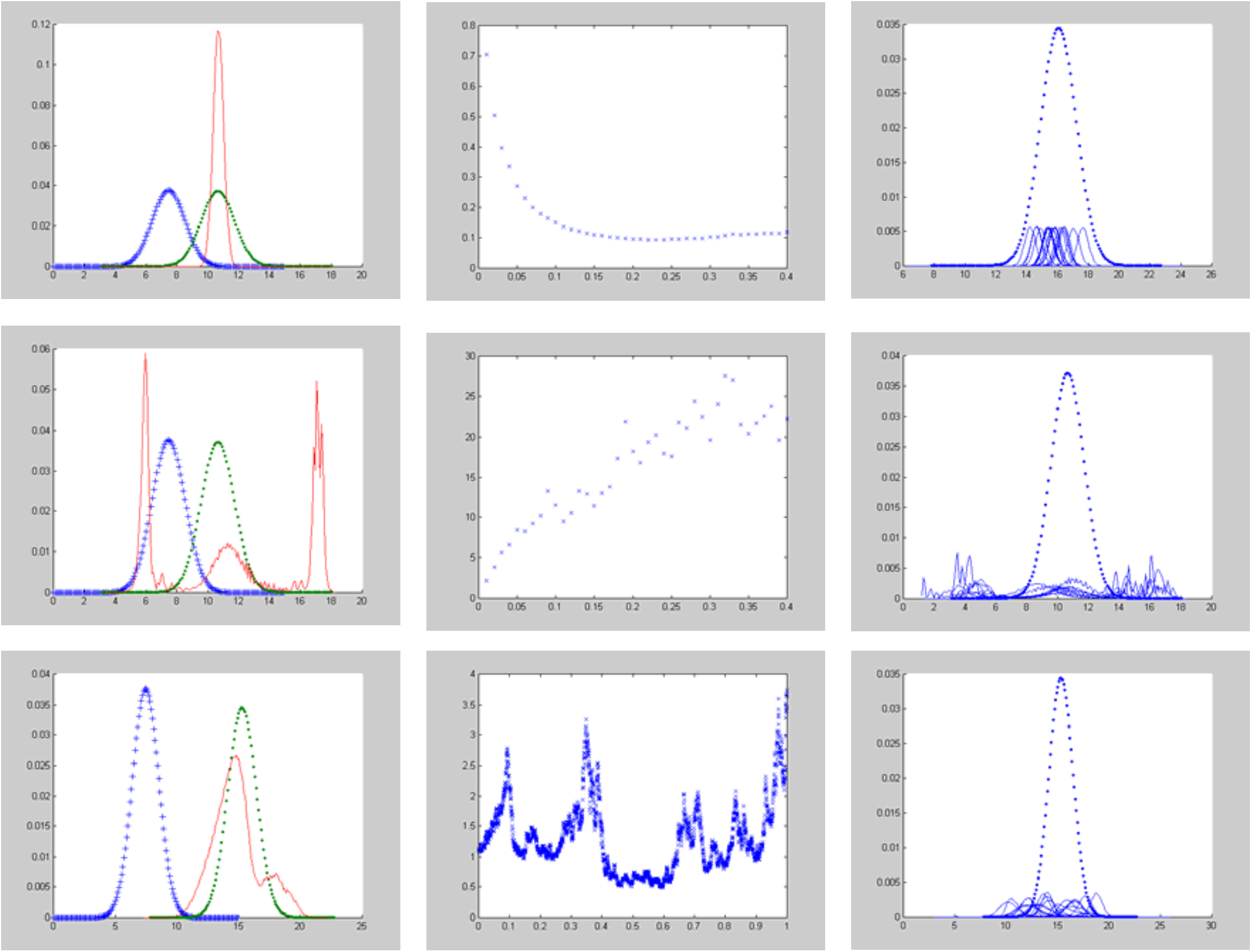

Fig. 6: Coherent and decoherent propagation of a wavepacket. First row, with localizing phase, φ=0, Second row: with delocalizing phase, φ=π, third row: with random phase φ. The first column displays, as in Fig. 1b, the initial wavepacket density (blue), after coherent evolution (green), and the ensemble average of the density after decoherent evolution. The second column displays, as in Fig. 2b, the mean square variation of position. The third column displays densities of individual wavefunctions after decoherent propagation, with their sum normalized to unity, and compared with the wavefunction after coherent propagation. Parameters are the same as in Figs 1-4.

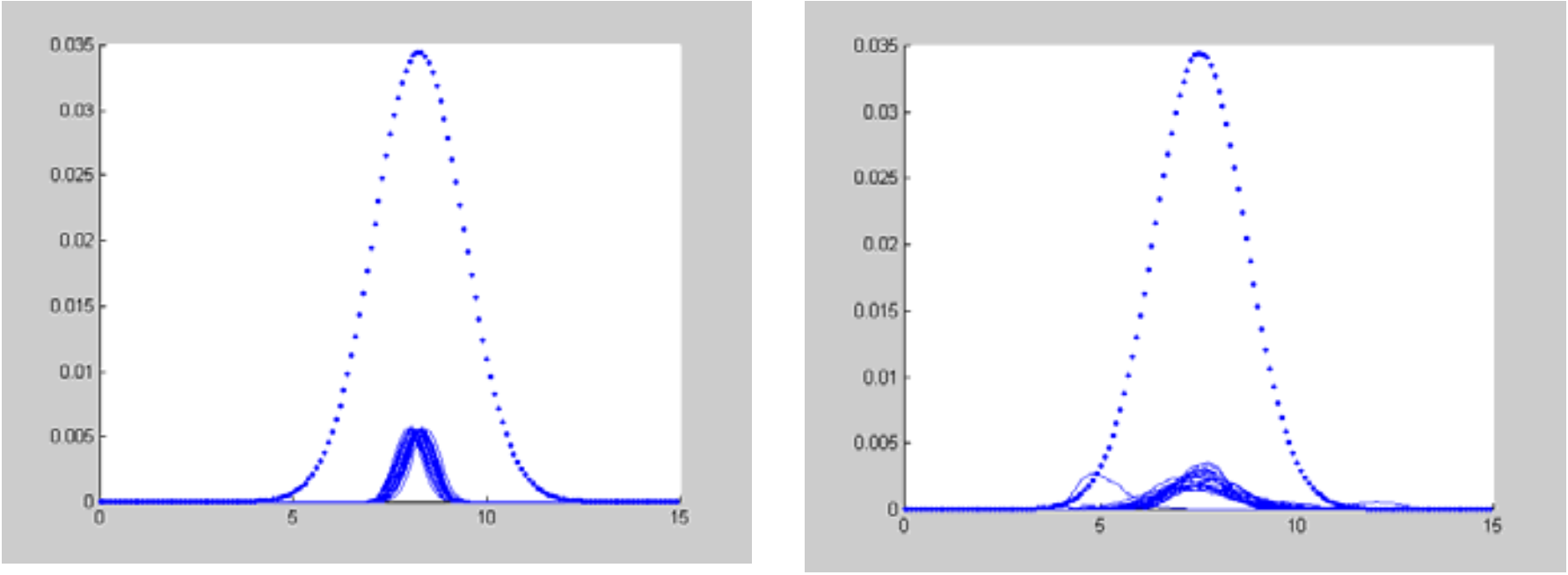

Fig. 7: centered individual decoherent wavefunction densities (a) localizing propagation φ=0, and (b) for neutral propagation φ random. In the case (b), the half-width of the decoherent and coherent wavefunctions are about the same.

positions of high amplitude will increase further by scattering for $\varphi = 0$, and decrease by scattering for $\varphi = \pi$. In the first case, the tendency towards concentration by scattering events and intrinsic spread of the wavefunction will cancel at a certain wavepacket width. In the second case, the intrinsic spread of the wavefunction is only enhanced by the scattering interaction, resulting in a complete delocalization of the wavepacket. This scenario is comfirmed by our simulations. In Fig. 6, the three essential cases are compared:

$\varphi = 0$;localizing propagation, first row in Fig. 6,

$\varphi = \pi$; delocalizing propagation, second row in Fig. 6,

$\varphi$ random; neutral propagation, third row in Fig. 6.

In the first case, the individual wavepacket width drops to a low, stable value, while in the second case, wavefunction delocalize. In the third case, there is apart from fluctuations little difference between coherent wavefunction density and individual decoherent wavefunction densities, other than the classical momentum spread. The last point is illustrated more clearly by Fig. 7.

**Relation to the plane-wave-decoherence model**

The phenomenological decoherence model based on scattering events after Eq.2 and localizing phase $\varphi=0$ yields common sense notions about the emergence of classical point particles from quantum theory. However Eq. 2 yet has to be related to a more fundamental theory. The basic idea of our phenomenological model is that the coherent propagation of a quantum particle is interrupted by scattering events. In a scattering event, the particle exchanges momentum with scattering field quanta which are described by plane waves. The system is open, with field quanta coming from infinity and leaving towards infinity. This situation appears to be very similar to the PWD model. In PWD, a single scattering event with momentum exchange $\kappa$, and scattering amplitude $\gamma$ transforms the initial state into the entangled state

$$\sqrt{1-\gamma}\left|\phi_{\mathbf{K}}\right\rangle\left|\chi_{\mathbf{k}}\right\rangle + e^{i\varphi}\sqrt{\gamma}\left|\phi_{\mathbf{K}+\kappa}\right\rangle\left|\chi_{\mathbf{k}-\kappa}\right\rangle . \tag{12}$$

Here, $\chi$ denotes field momentum eigenstates of the field, and the subscript the momentum of the field, or the central wavenumber of the wavepacket. Tracing out the field states, resulting in a mixture that can be described in terms of classical probabilities $p_1 = 1-\gamma$, and $p_2 = \gamma$ :

$$p_1\left|\phi_{\mathbf{K}}\right\rangle + p_2\left|\phi_{\mathbf{K}+\kappa}\right\rangle . \tag{13}$$

A scattering process (13) is however Eq. (2), with random phase $\varphi$, i. e. neutral propagation, Fig. 7, third row. The case of neutral propagation is equivalent to PWD.

Since the model obeying Eq. 2 with localizing phase cannot be realized by scattering of plane waves, we depart successively from PWD. In a first step, assume the field states $\chi$ in (12) to scatter at a second particle, similar to the first one, transferring the acquired momentum $-\boldsymbol{\kappa}$, returning to its original state **k**. Tracing out the field state now leaves us with an entangled two-particle state.

$$\sqrt{1-\gamma}\,|\phi_{\mathbf{K}}\rangle|\psi_{\mathbf{k}}\rangle+\sqrt{\gamma}\,|\phi_{\mathbf{K}-\kappa}\rangle|\psi_{\mathbf{k}+\kappa}\rangle\,. \tag{14}$$

By replacing (12) by (14) we have changed our model. Successive application of Eq. 12 describes a particle that becomes entangled with a growing number of field quanta. The successive application of (14) leads to a multiparticle state, i. e. to the quantum state of a gas of particles interacting by discrete scattering events.

In the recoil-free limit, $\gamma$ = ½ can be achieved by using the equivalence (5). For both particle states we now consider a new base, one base state being a term like in Eq. (2) with localizing phase $\varphi = 0$. Taken at the scattering position $\boldsymbol{x}_0$, the state is given by

$$\left|\phi_{L;\mathbf{K}-\boldsymbol{\kappa}/2}\right\rangle=\left|\phi_{\mathbf{K}-\boldsymbol{\kappa}}\right\rangle+\left|\phi_{\mathbf{K}}\right\rangle,\quad \left|\psi_{L;\mathbf{k}+\boldsymbol{\kappa}/2}\right\rangle=\left|\psi_{\mathbf{k}+\boldsymbol{\kappa}}\right\rangle+\left|\psi_{\mathbf{k}}\right\rangle, \tag{15}$$

while the state with delocalizing phase, $\varphi = \pi$ is

$$\left|\phi_{D;\mathbf{K}-\boldsymbol{\kappa}/2}\right\rangle=\left|\phi_{\mathbf{K}-\boldsymbol{\kappa}}\right\rangle-\left|\phi_{\mathbf{K}}\right\rangle,\quad \left|\psi_{D;\mathbf{k}+\boldsymbol{\kappa}/2}\right\rangle=\left|\psi_{\mathbf{k}+\boldsymbol{\kappa}}\right\rangle-\left|\psi_{\mathbf{k}}\right\rangle. \tag{16}$$

Expressing (14) in these terms yields

$$\left|\phi_{L;\mathbf{K}+\boldsymbol{\kappa}/2}\right\rangle\left|\psi_{L;\mathbf{k}-\boldsymbol{\kappa}/2}\right\rangle+\left|\phi_{D;\mathbf{K}+\boldsymbol{\kappa}/2}\right\rangle\left|\psi_{D;\mathbf{k}-\boldsymbol{\kappa}/2}\right\rangle. \tag{17}$$

In an absolute sense no particle localization results from subsequent processes of the kind of (14). However, particle states that lead to localization are entangled. *If* the second particle is localized, then the localization of the second particle follows from the model.

Successive application of scattering events according Eq. 17 leads to a multiparticle state whose Schmidt paths are partly classical, with localized particle states, and Schmidt paths composed of delocalized one-particle states, which have no obvious classical interpretation. In contrast, a multiparticle state where all terms of the Schmidt decomposition are “classical” has an obvious Many Worlds interpretation. In our example, that situation would be realized for a specific scattering process yielding a final state of localized wavefunctions only. Let *L1, L2* denote wavefunctions that are more localized than the corresponding one-particle wavefunction before scattering. Eq. 17 is then replaced by

$$\left|\phi_{L1;\mathbf{K}+\boldsymbol{\kappa}/2}\right\rangle\left|\psi_{L1;\mathbf{k}-\boldsymbol{\kappa}/2}\right\rangle+\left|\phi_{L2;\mathbf{K}+\boldsymbol{\kappa}/2}\right\rangle\left|\psi_{L2;\mathbf{k}-\boldsymbol{\kappa}/2}\right\rangle. \tag{18}$$

Subsequent application of (18) yields a multiparticle state in which every Schmidt component is classical, i.e. a product of localized single-particle wavepackets. The phenomenological model, applying (2) with zero phase is then well suited to describe the subjective experience on a randomly chosen Schmidt path, or “branch”.

Eq. (18) denotes a hypothetical scattering event allowing an unproblematic Many Worlds interpretation of the resulting multiparticle state. In the following, we investigate the entangled states that are actually produced by two-particle scattering, assuming a force-distance law. We choose a simplified situation: Initial and final states are composed of localized single-particle wavepackets. The scattering geometry is shown in Fig. 9, with the scattering parameter $q$ and unit vectors **e**, **e'**, and $\mathbf{e}_z$. In the following, we consider only the vertical dimension of the particle trajectories,

$$r \equiv \mathbf{r}(t)\cdot\mathbf{e}_z\,,\quad r' \equiv \mathbf{r}'(t)\cdot\mathbf{e}_z\,. \tag{19}$$

If the scattering parameter is large against the width of the wavepackets involved, $d << q$, an expansion of force-distance law is allowed. This situation has been analyzed in Refs. 6 and 20.

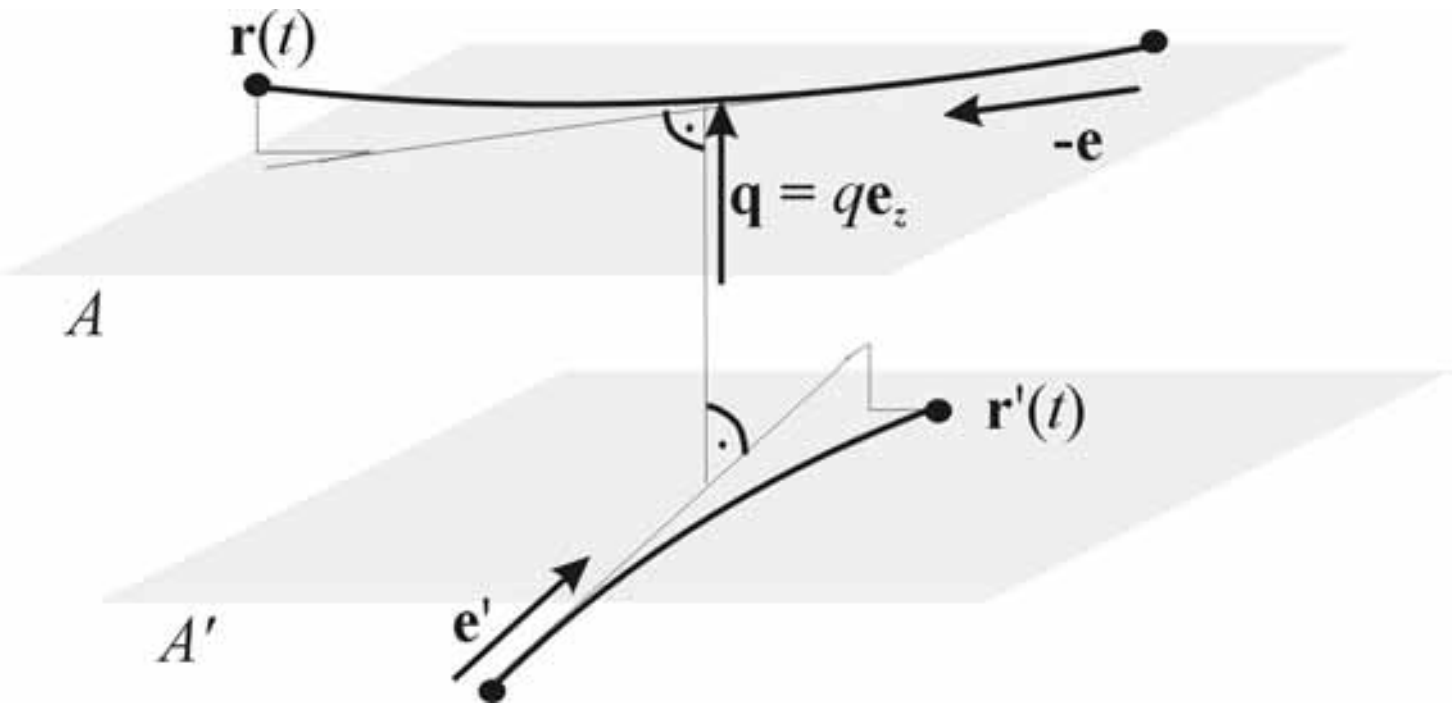

Fig. 8: Scattering between two particles

We model the each one-particle wavefunction by a pair of Gaussians wavepackets $\varphi(r,k,\sigma)$, whose centers are displayed as points in Fig.8.
The inital two-particle wavefunction is then given by

$$\psi_{\mathrm{ini}}(r,r') = \frac{1}{N}\left(\varphi(r-r_2,k)+\varphi(r-r_2,k)\right)\left(\varphi(r'-r_3,-k)+\varphi(r'-r_4,-k)\right), \tag{20}$$

with $N$ denoting a normalization factor, and the initial positions

$$r_1 = d+s,\; r_2 = -d+s,\; r_3 = d-\tilde{s},\; r_4 = -d-\tilde{s},\;\text{ and }\; \frac{2\pi}{k} << \sigma \approx d\,. \tag{21}$$

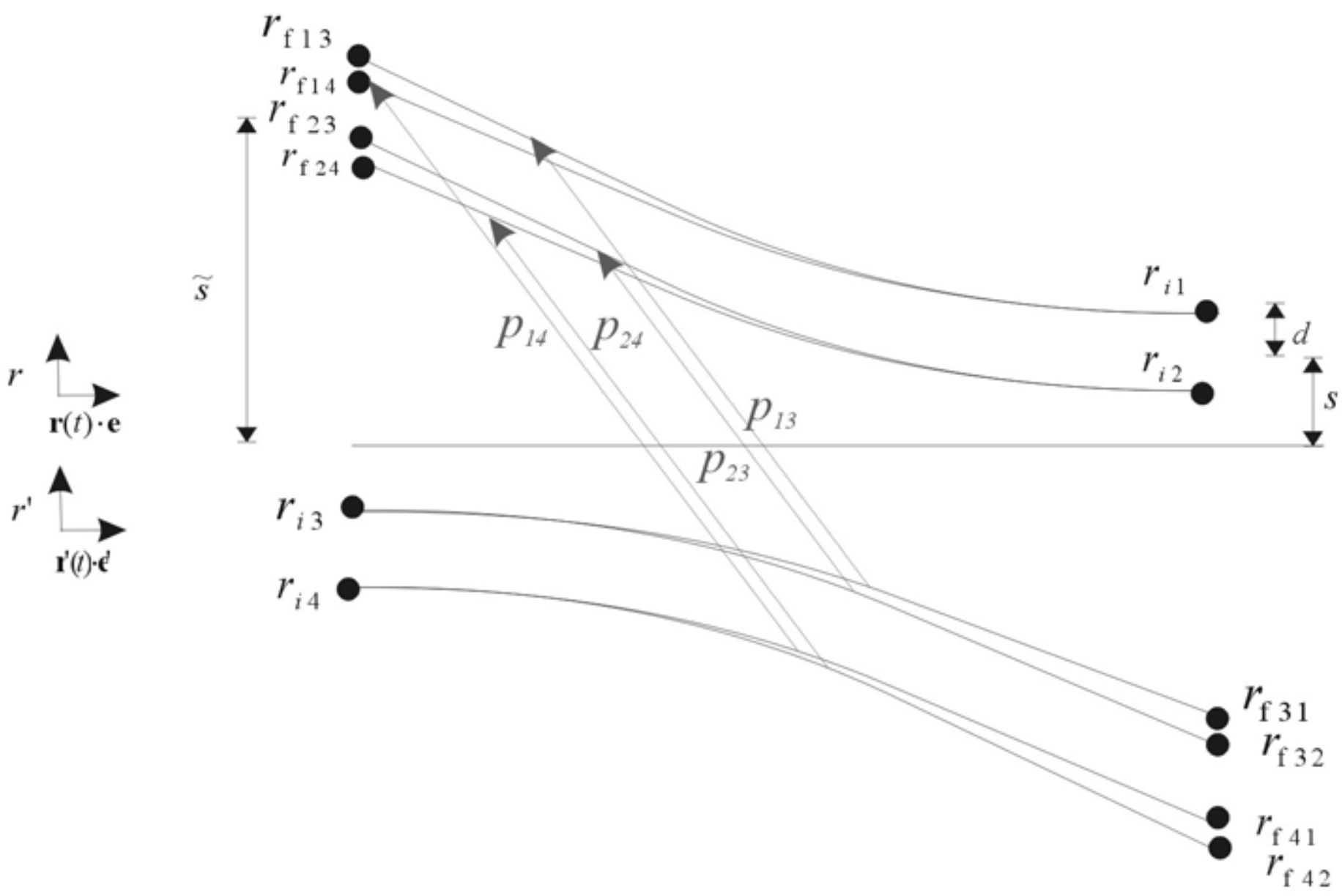

Fig. 9: Scattering between two particles with the initial state given by Eq. 20

After the scattering, wavepacket positions are given by

$$\begin{aligned} &r_{f13} = d + \tilde{s} + d_{13}, \quad r_{f14} = d + \tilde{s} + d_{14}, \quad r_{f23} = -d + \tilde{s} + d_{23}, \quad r_{f24} = -d + \tilde{s} + d_{24}, \\ &r_{f31} = d - s - d_{13}, \quad r_{f32} = d - s - d_{23}, \quad r_{f41} = -d - s - d_{14}, \quad r_{f42} = -d - s - d_{24}, \end{aligned} \tag{21}$$

The differences in the velocities resulting from the momentum transfer lead differences in position

$$d_{ij} = \delta v_{ij}\, t_c, \; \delta v_{ij} = \hbar k_{ij} / m. \tag{22}$$

with $t_c$ denoting the time between two scattering events. A scheme of the scattering process is shown in Fig.9.

We first consider the case $d_{ij} << d$, where the displacements of the wavepackets arising from the momentum transfer can be neglected. In Ref. 27, the Quantum Zeno regime is defined as an entangled state after a measurement-like interaction that has not separated in space. In this sense, our scattering event is of the Quantum Zeno type. The final state then be written as

$$\begin{aligned} \psi_f(r,r') = \frac{1}{N}\big(&\varphi(r - r_{1f}, k + k_{13})\varphi(r' - r_{3f}, -k - k_{13}) + \varphi(r - r_{1f}, k + k_{14})\varphi(r' - r_{4f}, -k - k_{14}) \\ &+ \varphi(r - r_{2f}, k + k_{23})\varphi(r' - r_{3f}, -k - k_{23}) + \varphi(r - r_{2f}, k + k_{24})\varphi(r' - r_{4f}, -k - k_{24})\big) \end{aligned} \tag{23}$$

with final positions $r_{1f} = d + \tilde{s}, r_{2f} = -d + \tilde{s}, r_{3f} = d - s, r_{4f} = -d - s,$ and the momentum transfers, which depend on the initial positions, Eq. (21): $p_{ij} = p(r_i, r_j) = -p_{ji} = \hbar k_{ij}$. This dependence $p(r_i, r_j)$ is expanded for small $d$: With the denotations

$$\tilde{k} = k + k_0, \quad \kappa = k_{\mathbf{0}} - k_{-1}, \; k_1 = \tilde{k} + \kappa, \tag{24}$$

Eq. 23 becomes

$$\psi_{\mathrm{f}}(r,r') = \frac{1}{N}\Big[\varphi\big(r - r_{1\mathrm{f}}, \tilde{k}\big)\varphi\big(r' - r_{3\mathrm{f}}, -\tilde{k}\big) + \varphi\big(r - r_{1\mathrm{f}}, k + \tilde{k} - \kappa\big)\varphi\big(r' - r_{4\mathrm{f}}, -(\tilde{k} - \kappa)\big) + \varphi\big(r - r_{2\mathrm{f}}, \tilde{k} + \kappa\big)\varphi\big(r' - r_{3\mathrm{f}}, -(\tilde{k} + \kappa)\big) + \varphi\big(r - r_{2\mathrm{f}}, \tilde{k}\big)\varphi\big(r' - r_{4\mathrm{f}}, -\tilde{k}\big)\Big] \tag{25}$$

For small momentum transfer κ, a WKB approximation can be made,

$$\psi_{\mathrm{f}}(r,r') = \frac{1}{N}\Big[\varphi\big(r - r_{1\mathrm{f}}, \tilde{k}\big)\varphi\big(r' - r_{3\mathrm{f}}, -\tilde{k}\big) + \varphi\big(r - r_{1\mathrm{f}}, k + \tilde{k}\big)e^{-i\kappa r}\varphi\big(r' - r_{4\mathrm{f}}, -\tilde{k}\big)e^{i\kappa r'} + \varphi\big(r - r_{2\mathrm{f}}, \tilde{k}\big)e^{i\kappa r}\varphi\big(r' - r_{3\mathrm{f}}, -\tilde{k}\big)e^{-i\kappa r'} + \varphi\big(r - r_{2\mathrm{f}}, \tilde{k}\big)\varphi\big(r' - r_{4\mathrm{f}}, -\tilde{k}\big)\Big] \tag{26}$$

Inserting the wavepacket positions yields

$$\psi_{\mathrm{f}}(r,r') = \frac{1}{N}\Big[\varphi\big(r - d + \tilde{s}, \tilde{k}\big)\varphi\big(r' - d - \tilde{s}, -\tilde{k}\big) + \varphi\big(r - d + \tilde{s}, \tilde{k}\big)\varphi\big(r' + d - \tilde{s}, -\tilde{k}\big)e^{-2i\kappa d} + \varphi\big(r + d + \tilde{s}, \tilde{k}\big)\varphi\big(r' - d - \tilde{s}, -\tilde{k}\big)e^{-2i\kappa d} + \varphi\big(r + d + \tilde{s}, \tilde{k}\big)\varphi\big(r' + d - \tilde{s}, -\tilde{k}\big)\Big]. \tag{27}$$

For a certain momentum transfer,

$$e^{-2i\kappa d} = i\,.$$

$$\psi_{\mathrm{f}}(r,r') = \frac{1}{N}\Big[\varphi\big(r - r_{1\mathrm{f}}, \tilde{k}\big)\varphi\big(r' - r_{3\mathrm{f}}, -\tilde{k}\big) + i\varphi\big(r' - r_{1\mathrm{f}}, \tilde{k}\big)\varphi\big(r' - r_{4\mathrm{f}}, -\tilde{k}\big) + i\varphi\big(r' - r_{2\mathrm{f}}, \tilde{k}\big)\varphi\big(r' - r_{3\mathrm{f}}, -\tilde{k}\big) + \varphi\big(r' - r_{2\mathrm{f}}, \tilde{k}\big)\varphi\big(r' - r_{4\mathrm{f}}, -\tilde{k}\big)\Big] \tag{28}$$

This two-particle state is displayed schematically in Fig. 10.

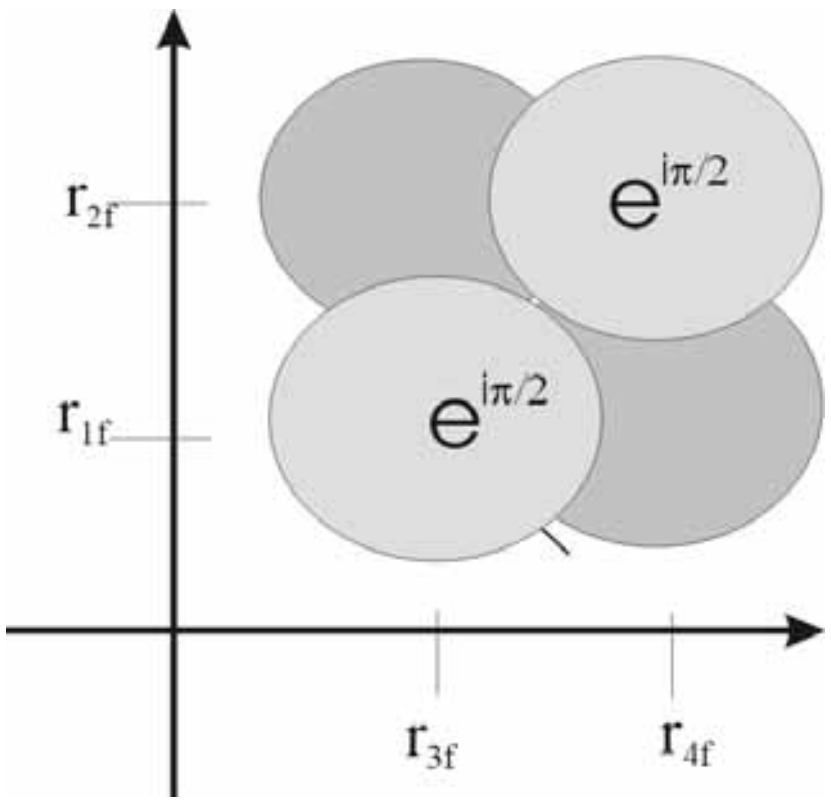

Fig. 10: Two-particle wavefunction, Eq. 28, of the final state after scattering with minimal momentum transfer

At this point, decoherence theory states that the phase of the singe-particle is no longer well-defined. It has to be described by a reduced density matrix. Beyond some decoherence length, interference is no longer possible.

Additional information is provided by writing the state (28) as a sum of two product states. For that purpose, we define the symmetric and antisymmetric superposition of the single-particle wavepackets

$$\phi^{+} = \varphi\big(r - r_{1\mathrm{f}}, \tilde{k}\big) + \varphi\big(r - r_{2\mathrm{f}}, \tilde{k}\big), \quad \phi^{-} = \varphi\big(r - r_{2\mathrm{f}}, \tilde{k}\big) - \varphi\big(r - r_{2\mathrm{f}}, \tilde{k}\big),$$
$$\phi'^{+} = \varphi\big(r' - r_{3\mathrm{f}}, \tilde{k}\big) + \varphi\big(r' - r_{4\mathrm{f}}, \tilde{k}\big), \quad \phi'^{-} = \varphi\big(r' - r_{3\mathrm{f}}, \tilde{k}\big) - \varphi\big(r' - r_{4\mathrm{f}}, \tilde{k}\big), \tag{29}$$

Note that the symmetic state is more localized than the antisymmetric state, assuming an overlap of the initial wavepackets.Expressed in that base, the two-particle state becomes

$$\psi_{\mathrm{f}}(r,r') = \frac{1}{N}\left[(1+i)\phi^{+}\phi'^{+} + (1-i)\phi^{-}\phi'^{-}\right] \tag{30}$$

The decomposition (2) into a localized and a delocalized oner-particle product state is shown schematically in Fig. 11.

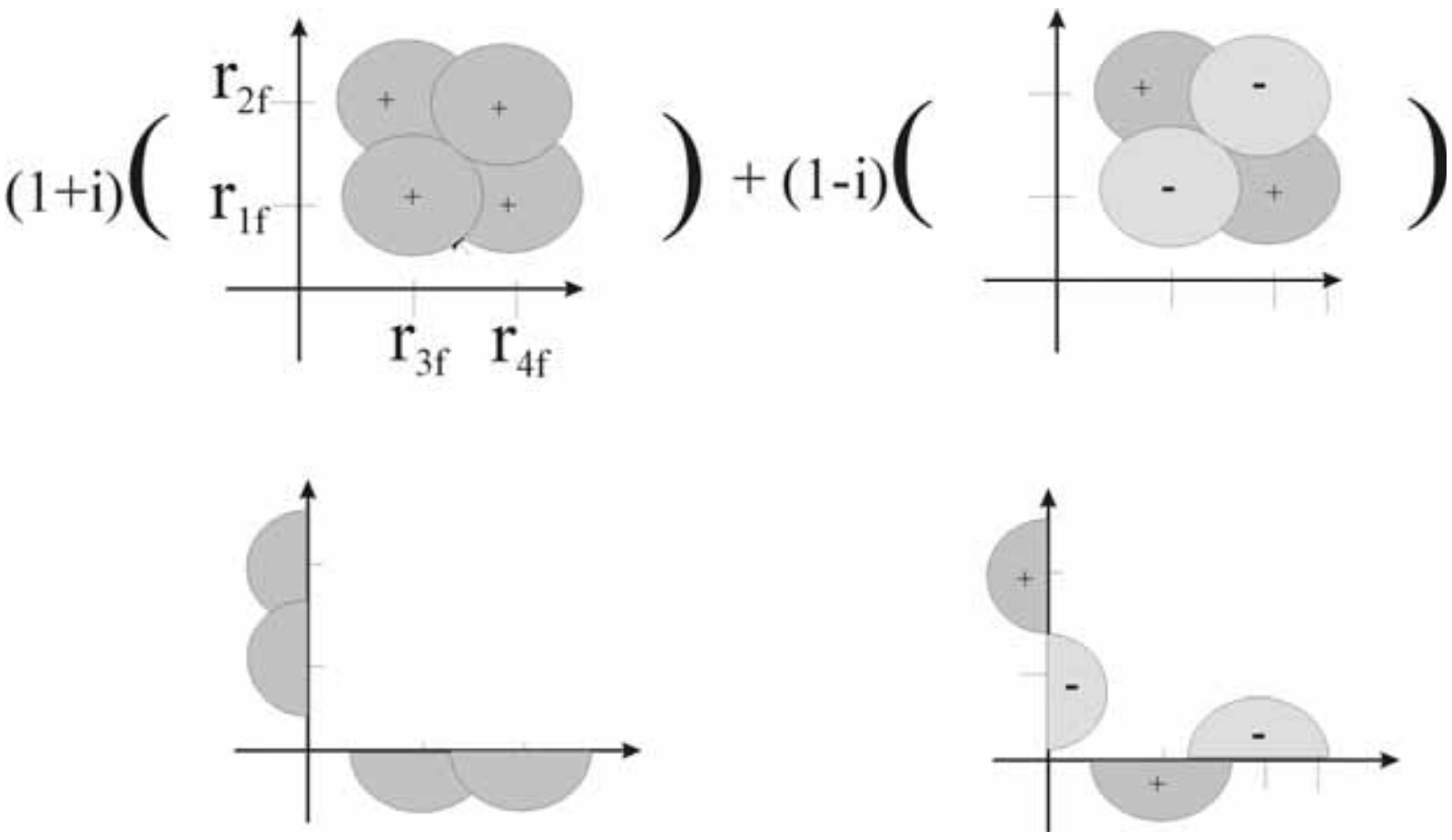

Fig. 11: The state of Fig. 10 written as a pair of one-particle product states

Note that the above base is also unique, when symmetry in the particle positions is desired.
If the initial state (20) is replaced by

$$\psi_{\mathrm{ini}}(r,r') = \frac{1}{N}\left(\varphi(r-r_{1\mathrm{i}},k) + i\varphi(r-r_{2\mathrm{i}},k)\right)\left(\varphi(r'-r_{3\mathrm{i}},-k) + i\varphi(r'-r_{4\mathrm{i}},-k)\right), \tag{31}$$

the localized and the delocalized states in (30) become

$$\begin{aligned}\phi^{+} &= \varphi(r-r_{1\mathrm{f}},\tilde{k}) + (\sqrt{2}-1)\varphi(r-r_{2\mathrm{f}},\tilde{k}), \quad \phi^{-} = \varphi(r-r_{2\mathrm{f}},\tilde{k}) - (\sqrt{2}-1)\varphi(r-r_{2\mathrm{f}},\tilde{k}),\\ \phi'^{+} &= \varphi(r'-r_{3\mathrm{f}},\tilde{k}) + (\sqrt{2}-1)\varphi(r'-r_{4\mathrm{f}},\tilde{k}), \quad \phi'^{-} = \varphi(r'-r_{3\mathrm{f}},\tilde{k}) - (\sqrt{2}-1)\varphi(r'-r_{4\mathrm{f}},\tilde{k}),\end{aligned} \tag{32}$$

In this case, the localized states $\phi^{+}$ or $\phi'^{+}$ are more localized than the initial state. In general, small-momentum scattering of two particles leads to a pair of entangled states. One-particle wavefunction of one Schmidt component are in general more localized than both of the one-particle wavefunction of one Schmidt component. This means, the scattering process obeys Eq. (17) , and therefore (2) with random phase. This is equivalent to PWD. Decoherence by itself does not yield localized wavepackets
We now turn to the case $d_{ij} \approx d$ . Fig. 12 shows the scattering event of Fig. 9 after a longer time interval, where the displacements related to the momentum transfers $p_{ij}$ become significant.

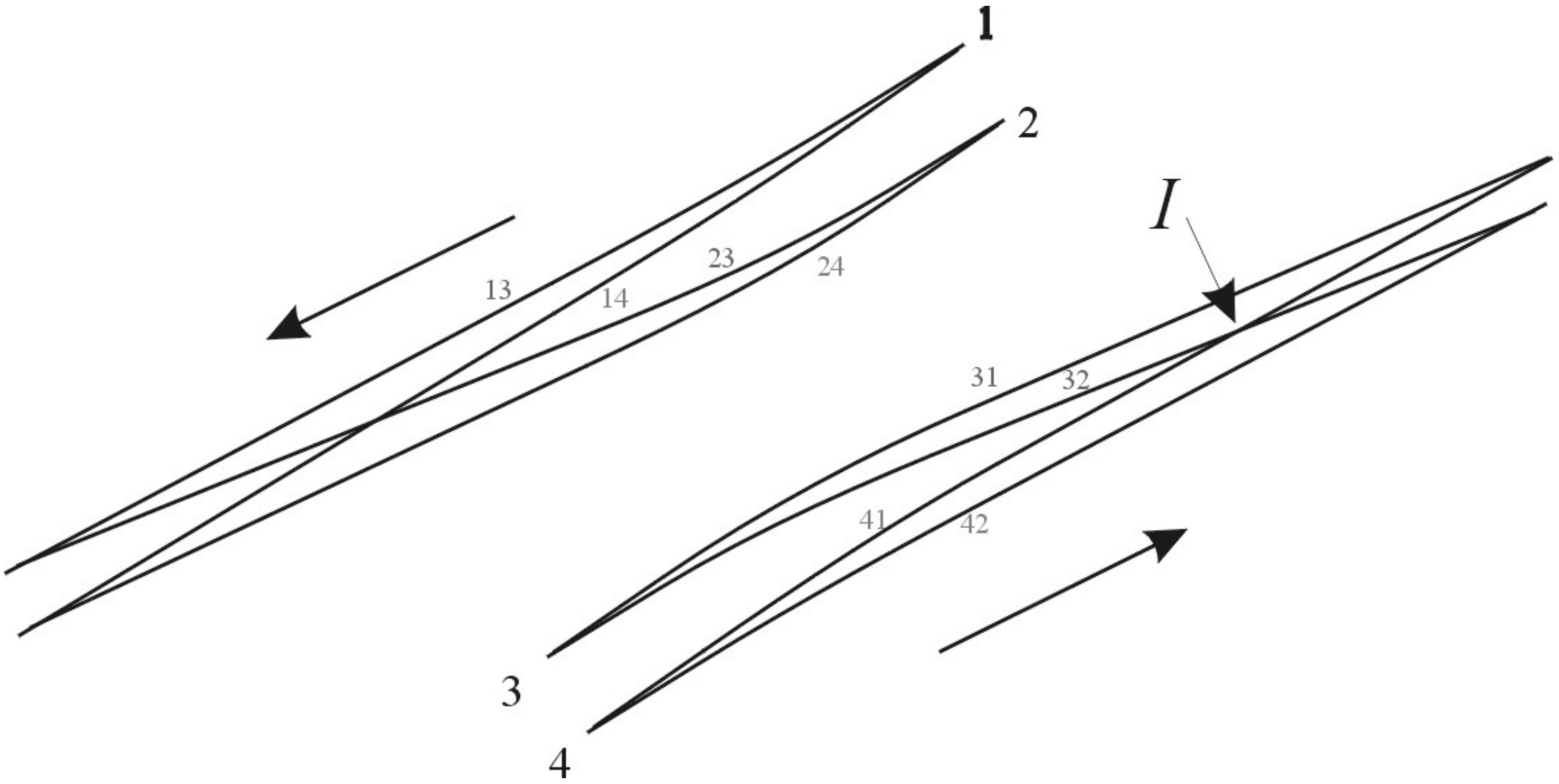

Fig. 12: Same as Fig. 9, but after a longer time. Trajectories of partial wavepackets intersect. The two-particle state at the intersection point I of trajectories 32 and 41 is analyzed further in the text.

After the scattering, particle trajectories intersect. This lensing effect of a repulsive potential is entirely classical and well-known. Before *t*, the width of the single-particle state shrinks. This means that for collision times of the order of *t*, a wavepacket become more effectively localized by scattering. In particular, we write down the first crossing point *I* from the scattering event. Here the trajectories 32 and 41 with extreme momentum transfers cross. The corresponding two-particle state is displayed in Fig. 13b. In Figs. 13c is shown that this state can be written in a good approximation as a pair of Schmidt components that are composed of one-particle wavepackets whose widths have decreased, compared with the one-particle states of the initial state. If the scattering particles have very different mass, momentum transfers are the same, and Figs. 10 and 13a would be unchanged. However the displacments are much larger for the lighter particle, so that 13b has to be replaced by 13d. Both particles become localized at the same rate. The effect of the momentum transfer on the path of the heavy particle are however negligible. Fig 13e is the analog of 13b for an attractive potential. Again, the multiparticle state can be understood as a pair of Schmidt components of localized single-particle wavefunctions. Therefore, for $d_{ij} << d$ , Eq. 18 can be assumed, and therefore Eq. 2 with φ=0.

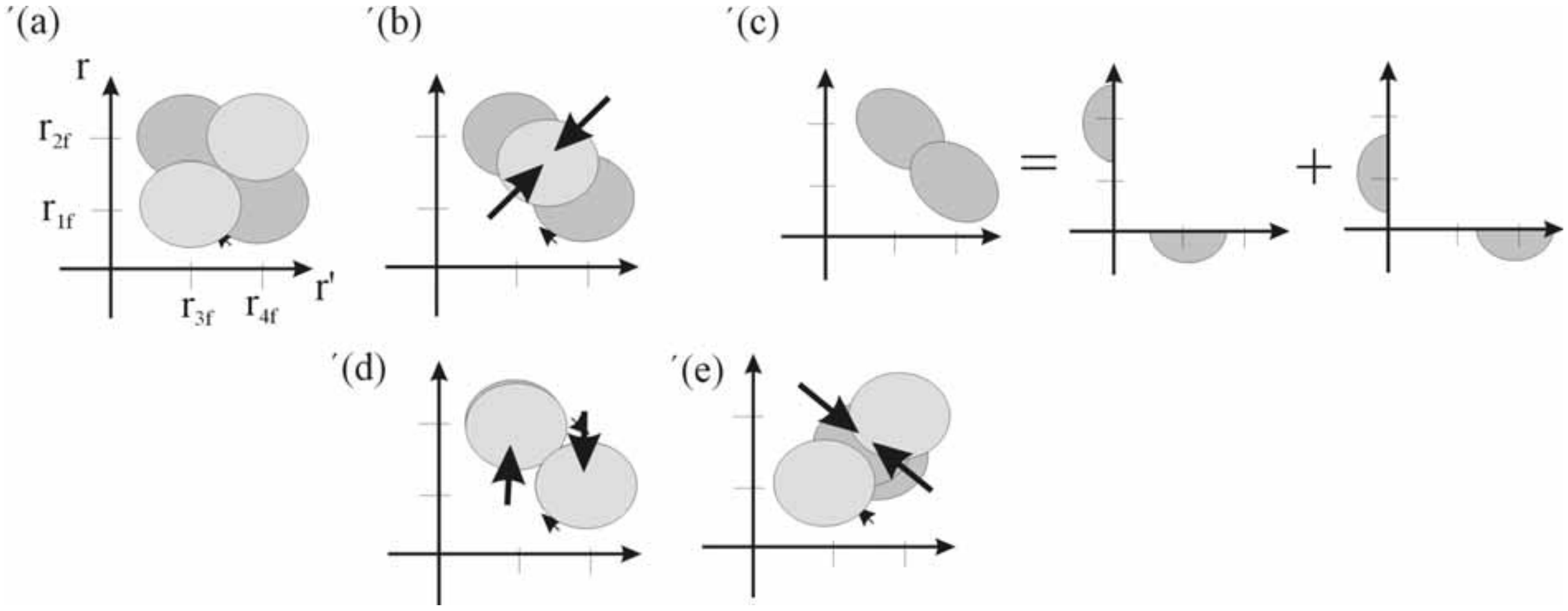

Fig. 13: Two-particle wavefunctions a) immediately after scattering, identical with Fig.11, b) at the intersection point I, c) The state b) written as a pair of products of one-particle states; d) two-particle state for particles of very different mass at the intersection point I of the lighter particle; e) two-particle state for particles of the same mass, but with attractive instead of repulsive interaction.

In the case $d_{ij} >> d$ , the multiparticle state has to be described by four Schmidt components in our simplified example, and by much more of them in general. We do not include this case into our model because of its complexity.

In conclusion, we have outlined a quantum model of a gas whose particles interact by scattering events with small momentum transfer. It has close relations to the model of decoherence in a gas in Ref. 6, and the more complicated toy model in Ref. 20. The model can be understood only within a Many Worlds interpretation using the Schmidt path approach. The model allows to track a single particle along a Schmidt path. In numerical simulation of standard problems, the model yields constant-width wavepackets and definite outcomes at barrier scattering. In this particular system, all Schmidt paths are classical. It therefore obeys the Born rule, avoiding a mayor conceptual difficulty of other relative state interpretations of quantum mechanics.

**Acknowledgement**
I thank Ulrich Nortmann and Daniel Schoch, Saarland University, for interesting discussions.